\documentclass[10pt]{article}

\usepackage{graphicx}%
\usepackage{amsmath,amssymb,amsthm,upref,bm}%
\usepackage{labelfig}%
\usepackage{mathrsfs}

\DeclareMathAlphabet{\varmathbb}{U}{pxsyb}{m}{n}

\newcommand{\D}{\mathrm{d}\kern0.2pt}%
\newcommand{\ii}{\kern0.05em\mathrm{i}\kern0.05em}%
\newcommand{\E}[1]{\textrm{e}^{#1}}%
\newcommand{\RR}{\varmathbb{R}}%

\begin{document}

\baselineskip=4.4mm

\makeatletter

\title{\bf On Delusive Nodes of Free Oscillations}

\author{Nikolay Kuznetsov}

\date{}

\maketitle

\vspace{-10mm}

\begin{center}
Laboratory for Mathematical Modelling of Wave Phenomena, \\ Institute for Problems
in Mechanical Engineering, Russian Academy of Sciences, \\ V.O., Bol'shoy pr. 61,
St. Petersburg 199178, Russian Federation \\ E-mail: nikolay.g.kuznetsov@gmail.com
\\[4mm] {\bf In memoriam of Vladimir Arnold}
\end{center}

\begin{abstract}
Two theorems and one conjecture about nodal sets of eigenfunctions arising in
various spectral problems for the Laplacian are reviewed. It occurred that all these
assertions are incorrect or only partly correct, but their analysis has brought
better understanding of the corresponding area of mathematical physics. The
contribution made by V.\,I. Arnold is emphasized.
\end{abstract}

\setcounter{equation}{0}

The name of Vladimir Arnold, who passed away on 3 June 2010, is well known to
mathematicians all over the world. Indeed, along with the Kolmogorov--Arnold--Moser
theory about the stability of integrable systems (his best known contribution to
mathematics) there are several other notions associated with him, for example: the
Arnold conjecture about the number of fixed points that a smooth function has on a
closed manifold, Arnold's cat map, the Arnold diffusion, an Arnold tongue in
dynamical systems theory.

A biographical sketch of Vladimir Igorevich Arnold by O'Connor and Robertson (an
entry of MacTutor History of Mathematics archive) is available online at
http://www-history.mcs.st-andrews.ac.uk/Biographies/Arnold.html. A lot of
interesting details about Arnold's life and work are presented by his colleagued and
disciples in the tribute and memories published in 2012; see \cite{Not1} and
\cite{Not2}, respectively. From these notes one gets a clear idea that everybody,
who maintained contact with him, was greatly impressed by his extraordinary
personality.

Among Arnold's numerous honours one finds Dannie Heineman Prize for Mathematical
Physics awarded in 2001 jointly by the American Physical Society and American
Institute of Physics. This is not an incident because he had a deep feeling of the
unity of mathematics and natural sciences. His often quoted remark says that
mathematics is a part of physics in which experiments are cheap.

\footnotetext{An item in the collection dedicated to the 75th anniversary of the
Steklov Mathematical Institute in Moscow. Before 1934, when the Soviet Academy of
Sciences was moved from Leningrad to Moscow, this institute was a division of the
Physical--Mathematical Institute organised by V.\,A.~Steklov in 1921 (see Steklov's
recollections cited in \cite{K}).}

No wonder that one of Arnold's papers published posthumously deals with an
important property of eigenoscillations in Mathematical Physics; see \cite{A},
submitted for publication six months before his death. In this paper$^0$, Arnold
with his inherent mastery of both the subject and storytelling, describes a
fascinating fact that an incorrect theorem was announced in the classical book
\cite{CHE} by Courant and Hilbert. (This edition is cited in \cite{A}, but, for the
reason explained below, Arnold used either the 2nd German edition \cite{CHD} or,
most likely, its Russian translation published twice in 1933 and 1951.)

The theorem in question deals with nodal sets (or, for brevity, nodes) of linear
combinations of some particular eigenfunctions (see the next paragraph). Such a set is
simply defined as the set, where a function vanishes. To make the importance of
eigenfunctions clear, we just mention that they serve for describing free
oscillations of strings and membranes and nodes show, where an oscillating object is
immovable because, by its definition a node separates the sets, where the function
is positive and negative. In one, two and three dimensions, nodal sets consist of
points, curves and surfaces, respectively. Pictures of nodal curves for some modes
of oscillations of the square membrane fixed along its boundary can be found in many
textbooks (see, for example, \cite{Str}, p.~266).

\begin{figure}[t]
\centering \mbox{\kern-2mm}\includegraphics[width=50mm]{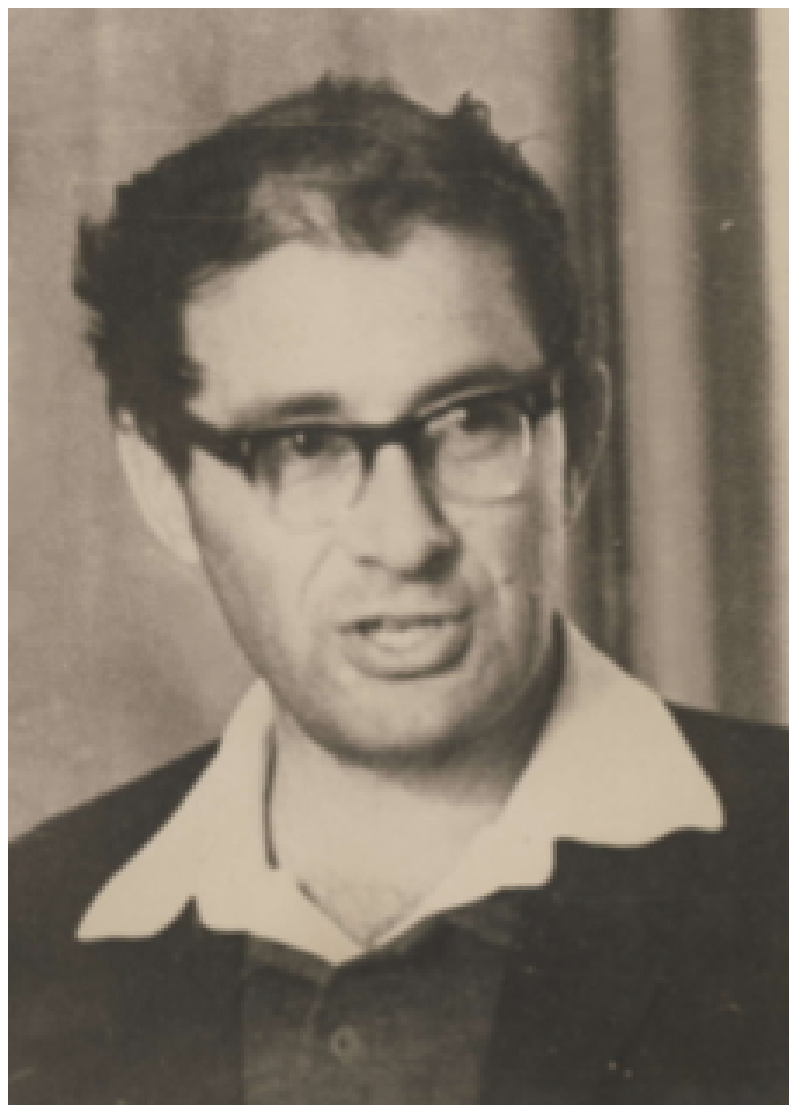} \\[3mm] Vladimir
Igorevich Arnold in 1977. \vspace{-4mm}
\end{figure}

It is amazing that there are rather many theorems and conjectures proved to be
incorrect in this area of research. Let us list those considered in this paper and
recall other renowned questions concerning the same spectral problems of
mathematical physics. We begin with the theorem which is the topic of the Arnold's
paper \cite{A}. It concerns nodes of linear combinations of eigenfunctions of the
Dirichlet Laplacian and we illustrate the question's essence with some elementary
examples. This material is presented in the first section.

What is widely known about the eigenvalue problem for the Dirichlet Laplacian is the
question `Can one hear the shape of a drum?' posed by Mark Kac in 1966 in the title
of his paper \cite{MK}. However, this question is about the whole set of
eigenvalues, whereas there are many subtle questions about properties of
eigenfunctions corresponding to individual eigenvalues. One of them, referred to as
Payne's conjecture, concerns nodes of the second eigenfunction; being more
technical, it is considered in the third section.

It is worth mentioning that the negative answer to the Kac's question was obtained
in 1992; it is presented in the form accessible to a general audience in the article
\cite{GW}. However, this answer, like falsity of the above mentioned theorem
discussed in \cite{A}, is only a part of the story. In November 2012, S.~Titarenko
presented another part at the Smirnov Seminar on Mathematical Physics in
St.~Petersburg (http://www.pdmi.ras.ru/$\sim$matfizik/seminar 2012-2013.htm). The
most important point of his talk entitled `When can one hear the shape of a drum?
Sufficient conditions' is that for answering in the positive to the Kac's question
the boundary of drum's membrane must be smooth. Indeed, smoothness is violated in
all of the now numerous examples delivering the negative answer (see, for example,
\cite{GT}, p.~2235; this article also contains an extensive list of references on
mathematical and physical aspects of isospectrality). Unfortunately, Titarenko's
result is still unpublished.

The second section deals with the well known phenomenon of liquid sloshing in
containers (widely used examples of these are tea cups, coffee mugs, vine glasses,
cognac snifters {\it etc}). The corresponding mathematical mo\-del\,---\,the so-called
sloshing problem (it is also referred to as the mixed Steklov
problem)\,---\,attracted much attention after awarding the 2012 Ig~Nobel Prize for
Fluid Dynamics to R.~Krechetnikov and H.~Mayer for their investigation why coffee so
often spills while people walk with a filled mug \cite{KM2012}. This effect results
from the correlation between the fundamental sloshing frequency and that of steps.
Here, a property of the sloshing nodes (the liquid remains immovable there during
its free oscillations) is considered. The presented example demonstrates that a gap
in the proof of a certain theorem describing the behaviour of nodes cannot be filled
up.

Another aim of this paper is to show how application of rather simple tools (in
particular, an analysis of the behaviour of functions defined explicitly, for
example, by improper integrals and even by elementary trigonometrical formulae)
leads to interesting results concerning important questions that challenge both
mathematical and physical intuition. It should be emphasised that such questions
were among Arnold's favorites. Indeed, his unique intuition, for example, in the
subject of catastrophes allowed him to guess on the spot the right answers when
physicists and engineers asked him what kind of catastrophic effect could be
expected in their problems. Many of his guesses were based on very simple models
like that considered in the next section.

\subsection*{Arnold on a footnote in the Courant--Hilbert book}

Arnold begins his story with the following \\[-6mm]
\begin{quote}
topological result [\dots] valid on any compact ma\-nifold: an eigenfunction $u$ of
the Laplace operator \\[-2mm]
\[ \Delta u = \lambda u \ \ \mbox{with eigenvalue} \ \lambda = \lambda_n
\]
(we arrange them in order of increasing frequencies $- \lambda_1 \leq - \lambda_2
\leq - \lambda_3 \leq \dots$) vanishes on the oscillating manifold $M$ in a way such
that its zeros divide $M$ into at most $n$ parts.
\end{quote}
In its original form, the result obtained by Courant in 1923 concerns nodes of
eigenfunctions of a self-adjoint second order differential operator (for example,
the Sturm--Liouville operator on an interval and the Laplacian in a bounded
higher-dimensional domain) with one of the standard boundary conditions (for
example, the Dirichlet and Neumann conditions). Namely, Courant's theorem asserts
that (see \cite{CHE}, p.~452): \\[-4mm]
\begin{quote}
if [the] eigenfunctions are ordered according to increasing eigenvalues, then the
nodes of the $n$th eigenfunction divide the domain into no more than $n$ subdomains.
No assumptions are made about the number of independent variables.
\end{quote}

\begin{figure}[t]
\vspace{1.6mm} \centering \mbox{\kern-2mm}\includegraphics[width=60mm]{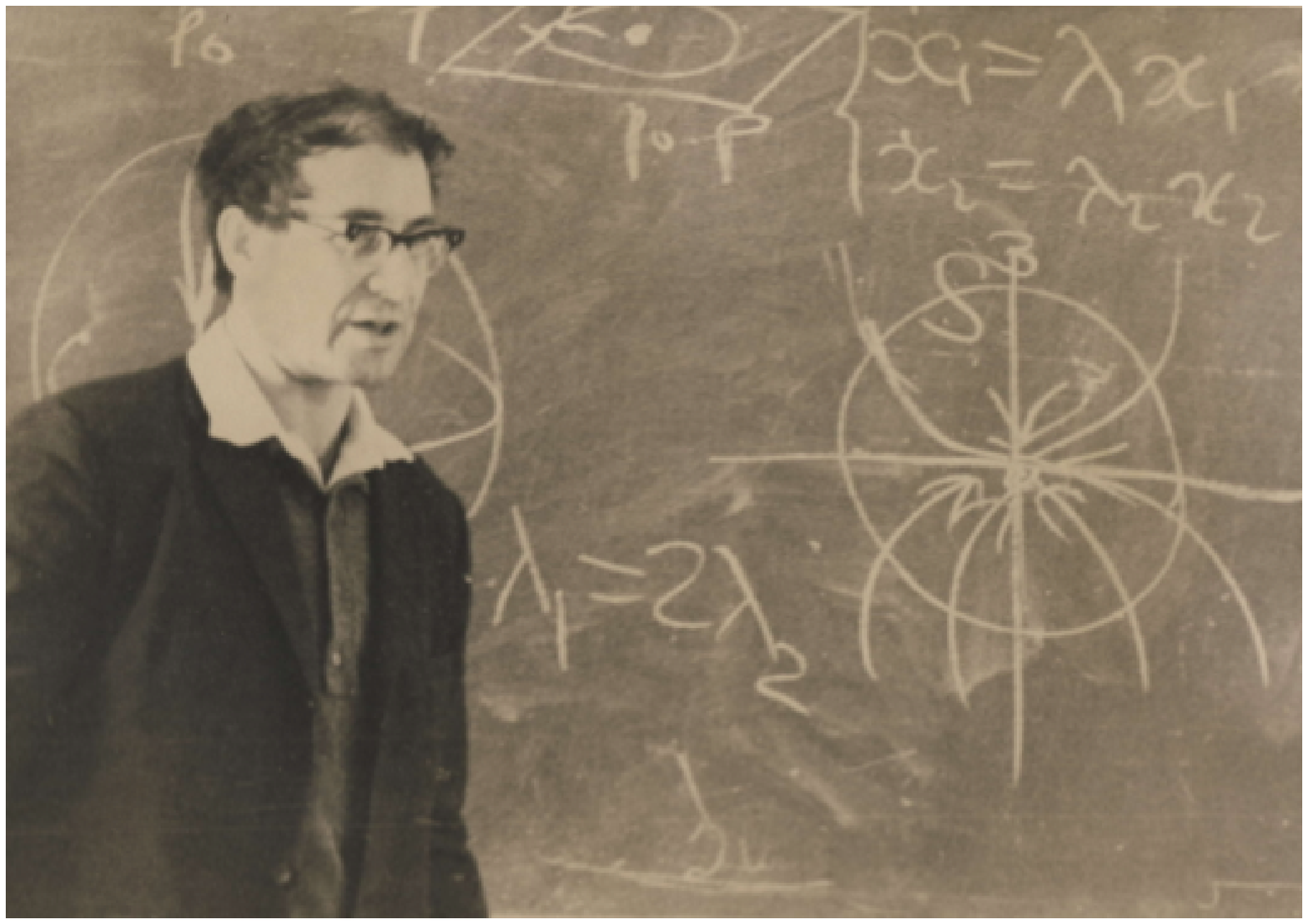}
\\[2mm] V. Arnold lecturing in Syktyvkar in 1977. \vspace{-4mm}
\end{figure}

Two simplest examples illustrating this theorem are delivered by the equation
describing the set of possible shapes of an homogeneous string in free time-harmonic
oscillations: \\[-3mm]
\begin{equation}
- u'' = \lambda u \quad \mbox{on} \ (0, \pi) , \label{1}
\end{equation}
augmented by either the Dirichlet conditions \\[-2mm]
\begin{equation}
u (0) = u (\pi) = 0 , \label{2}
\end{equation}
which means that the ends of a string are fixed, or the Neumann conditions \\[-2mm]
\begin{equation}
u' (0) = u' (\pi) = 0  \label{3}
\end{equation}
when the ends are free. It is clear that the eigenfunction $u_n = \sin n x$,
$n=1,2,\dots$, corresponds to $\lambda_n = n^2$ under the boundary conditions
(\ref{2}), whereas conditions (\ref{3}) give \\[-2mm]
\[ u_n = \cos (n-1) x \ \ \mbox{and} \ \lambda_n = (n-1)^2 , \ \mbox{respectively}.
\]
Note that in both cases the $n$th eigenfunction divides the interval into precisely
$n$ parts. Courant proves that this property remains valid for a general
Sturm--Liouville problem.

Prior to proving the latter result, a footnote announcing the notorious incorrect
theorem appears at the end of the proof of the theorem cited above (see the first
footnote on p. 454 in \cite{CHE}): \\[-6mm]
\begin{quote}
The theorem just proved may be generalized as follows: Any linear combination of the
first $n$ eigenfunctions divides the domain, by means of its nodes, into no more
than $n$ subdomains. See the G\"ottingen dissertation of H. Herrmann, Beitr\"age zur
Theorie der Eigenwerte und Eigenfunctionen, 1932.
\end{quote}
Below, this assertion is referred to as Herrmann's theorem. Arnold writes about it:
\\[-6mm]
\begin{quote}
This {\it generalization of Courant’s theorem} is not proved at all in the book by
Courant and Hilbert; it was just mentioned that the proof ``will soon be published
(by a disciple of Courant)''.
\end{quote}
From the last sentence we see that Arnold used either the 2nd German edition
\cite{CHD} published in 1931 or, what is more likely, its Russian translation. Then
he continues: \\[-6mm]
\begin{quote}
Having read all this, I wrote a letter to Courant, ``Where can I find this proof
now, 40 years after Courant announced the theorem?'' Courant answered that ``one can
never trust one’s students: to any question they answer either that the problem is
too easy to waste time on, or that it is beyond their weak powers''.

As regards Courant and Hilbert's {\it Mathematical Physics}, according to Courant's
published recollections, this book was nevertheless written by his students.
\end{quote}
Of course, Arnold exaggerates the role of students, but at the beginning of preface
to \cite{CHE} Courant writes that the second German edition was ``revised and
improved with the help of K.\,O. Friedrichs, R. Luneburg, F. Rellich, and other
unselfish friends''.

Soon after receiving Courant's reply, Arnold discovered that applying Herrmann's
theorem to the eigenfunctions of the Laplacian on the sphere $S^N$ with the standard
Riemannian metric one obtains an estimate for the number of components complementing
a real algebraic hypersur\-face of the degree $n$ in the $N$-dimensional projective
space (see \cite{A1}). The idea behind this is that the so-called spherical
harmonics (eigenfunctions of the Laplacian on the two-dimensional sphere) are
defined as follows. The set of these functions corresponding to the $n$th eigenvalue
consists of traces on $S^2$ of homogeneous harmonic polynomials of the degree $n-1$
in $\RR^3$ (see \cite{Str}, p.~263). Hence a linear combination of eigenfunctions
corresponding to the first $n$ eigenvalues is also a harmonic polynomial whose
degree is bounded by $n$. In \cite{A}, Arnold comments his estimate as follows:
\\[-6mm]
\begin{quote}
[\dots] it turned out that the results of the topology of algebraic curves that I
had derived from the generalized Courant theorem contradict the results of quantum
field theory. Nevertheless, I knew that both my results and the results of quantum
field theory were true. Hence, the statement of the generalized Courant theorem is
not true (explicit counterexamples were soon produced by Viro). Courant died in 1972
and could not have known about this counterexample.
\end{quote}
Indeed, seven years after Courant's death, Viro found an example of real algebraic
hypersurface for which Arnold's estimate does not hold, thus establishing what is
incorrect about Herrmann's theorem. Namely, it is valid only under some restrictions
on the number of independent variables, in particular, it is false for the Laplacian
on $S^3$ and higher-dimensional spheres (see \cite{V}).

However, Herrmann's theorem is true for eigenfunctions of the Dirichlet and Neumann
problems for equation (\ref{1}). This follows from elementary trigonometric formulae
(see 1.331.1 and 1331.3 in \cite{GR}). Indeed, if $n > 1$, then the $n$th Dirichlet and
Neumann eigenfunctions can be written as follows: \\[-4mm]
\begin{eqnarray}
&& \!\!\!\!\!\!\!\!\!\!\!\!\!\!\!\!\!\!\!\!\!\!\!\!\!\!\!\! \sin n x = \sin x
\sum_{k=0}^{[ (n-1)/2 ]} (-1)^k \left( \!\!\!\!\!\! \begin{array}{c} \ n - k - 1 \
\\ k \end{array} \!\!\!\!\!\! \right) (2 \cos x)^{n - (2k+1)} , \label{4} \\ &&
\!\!\!\!\!\!\!\!\!\!\!\!\!\!\!\!\!\!\!\!\!\! \cos (n-1) x = 2^{n-2} \cos^{n-1} x
\nonumber \\ && \!\!\!\!\!\!\!\!\!\!\!\!\!\! + \frac{n-1}{2} \sum_{k=1}^{[ (n-1)/2
]} \frac{(-1)^k}{k} \left( \!\!\!\!\!\! \begin{array}{c} \ n - k - 2 \ \\ k -1
\end{array} \!\!\!\!\!\! \right) (2 \cos x)^{n - (2k+1)} . \label{5}
\end{eqnarray}
Here $[m]$ stands for the integer part of $m$.

According to formula (\ref{4}), a linear combination of the first $n$ Dirichlet
eigenfunctions is the product of $\sin x$ and a polynomial of $\cos x$ whose degree
is at most $n-1$. Therefore, it has at most $n-1$ zeros and the number of nodes on
$(0, \pi)$ is also less than or equal to $n-1$. The similar conclusion follows from
(\ref{5}) for a linear combination of the first $n$ Neumann eigenfunctions. Let us
illustrate this considering linear combinations of the first two Dirichlet and
Neumann eigenfunctions which are \\[-2mm]
\[ \sin x (C_1 + 2 C_2 \cos x) \ \ \mbox{and} \ \ C_1 + C_2 \cos x , \ 
\ \mbox{respectively} .
\]
Here $C_1$ and $C_2$ are some constants. Both linear combinations have at most one
node on $(0, \pi)$. It exists when $C_2 \neq 0$ and also \\[-2mm]
\[ \left| \frac{C_1}{C_2} \right| < 2 \ \ \mbox{and} \ \ \left| \frac{C_1}{C_2} 
\right| < 1 
\] 
for the combinations of the Dirichlet and Neumann eigenfunctions, respectively.
These conditions are also necessary for the existence of a node.

In the same way, one obtains that Herrmann's theorem is true for eigenfunctions of
the following problem: \\[-2mm]
\[ - u'' = \lambda u \ \ \mbox{on} \ (0, 2 \pi) , \quad u (0) = u (2 \pi) , \quad 
u' (0) = u' (2 \pi) .
\]
For this periodic boundary value problem we have:

\vspace{1mm}

\noindent $\bullet$ The 1st eigenvalue is zero; it is simple and the corresponding
eigenfunction is a non-zero constant.

\noindent $\bullet$ The $n$th eigenvalue is $(n-1)^2$; its multiplicity is two and
the corresponding eigenfunctions are $\sin (n-1) x$ and $\cos (n-1) x$.

\vspace{1mm}

\noindent Then Herrmann's theorem is again a consequence of formulae (\ref{4}) and
(\ref{5}), but $n$ must be changed to $n-1$ in (\ref{4}).

In the second section of \cite{A1}, Arnold turns to the following Sturm--Liouville
problem: \\[-2mm]
\begin{equation}
- u'' + q u = \lambda u \ \ \mbox{on} \ (0, \ell) , \quad u (0) = u (\ell) = 0 ,
\label{SL}
\end{equation}
here $q$ is a positive function on $[0, \ell]$. He outlines Gel'fand's idea how to
prove Herrmann's theorem for eigenfunctions of this problem. It consists in
replacing \\[-6mm]
\begin{quote}
the analysis of the system of $n$ eigenfunctions of the one-particle
quantum-mechanical problem by the analysis of the first eigenfunction of the
$n$-particle problem (considering, as particles, fermions rather than bosons).
\end{quote}

\vspace{-2mm}

This approach so attracted Arnold that he included Herr\-mann's theorem for
eigenfunctions of problem (\ref{SL}) together with Gel'fand's hint into the 3rd
Russian edition of his {\it Ordinary Differential Equations} (see Problem~9 on the
list of supplementary problems at the end of \cite{A2}).

In \cite{A} Arnold devotes two pages to some details of Gel'fand's analysis, but at
the end he writes. \\[-5mm]
\begin{quote}
Unfortunately, the arguments above do not yet provide a proof for this generalized
theorem: many facts are still to be proved. [\dots]

Gel'fand did not publish anything concerning this: he only told me that he hoped his
students would correct [\dots] his theory. He pinned high hopes on V.\,B.~Lidskii
and A.\,G.~Kostyuchenko. Viktor Borisovich Lidskii told me that ``he knows how to
prove all this.'' [\dots] Although [his] arguments look convincing, the lack of a
published formal text with a rigorous proof of the Courant--Gel'fand theorem is
still distressing.
\end{quote}

\vspace{-1mm}

This is still true, but there is a hope that Victor Kleptsyn (Institut de Recherche
Math\'ematique de Rennes) will soon fill in this gap. In his unpublished manuscript,
he not only provides a proof for all gaps remaining in the above approach, but also
suggests an alternative one using the heat equation.

\subsection*{On sloshing nodal curves}

A particular case of the mixed Steklov eigenvalue problem gives the so-called
sloshing frequencies and the corresponding wave modes, that is, the natural
frequencies and modes of the free motion of water occupying a reservoir. When the
latter is an infinitely long canal of uniform cross-section $W$, the two-dimensional
problem arises. In this case, the boundary $\partial W$ consists of $F = \{ |x| <
a,\ y = 0 \}$ and $B = \partial W \setminus \bar F$ lying in the half-plane $y<0$.
The former is referred to as the {\it free surface} of water, whereas the latter is
{\it canal's bottom}.

The velocity potential $u (x,y)$ with the time-harmonic factor removed must satisfy
the following boundary value problem: \\[-6mm]
\begin{eqnarray}
&& u_{xx} + u_{yy} = 0\quad {\rm in} \ W, \label{lap} \\ && u_y = \lambda u \quad {\rm
on}\ F, \label{nu} \\ && \frac{\partial u}{\partial {\bf n}} = 0\quad {\rm on} \ B
. \label{nc}
\end{eqnarray}
Here {\bf n} denotes the exterior unit normal on $B$ and $\lambda = \omega^2/g$ is
the spectral parameter to be found along with $u$ ($\omega$ is the radian frequency
of the water oscillations and $g$ is the acceleration due to gravity). In order to
exclude the non-physical zero eigenvalue of (\ref{lap})--(\ref{nc}), it is usual to
augment the problem's statement by the orthogonality condition \\[-2mm]
\begin{equation}
\int_F u \, \D x = 0 . \label{ort}
\end{equation}

The condition on $F$ is the Steklov boundary condition first introduced by Steklov
in 1896, but the standard reference for the Steklov problem is the paper \cite{S}
published in 1902. Problem (\ref{lap})--(\ref{ort}) and its three-dimensional
version has been the subject of a great number of studies over more than two
centuries; see \cite{FK} for a historical review, whereas early results are
presented in the Lamb's classical treatise {\it Hydrodynamics} \cite{L}.

It is well-known that this problem has a discrete spectrum, that is, an infinitely
increasing sequence of positive eigenvalues of finite multiplicity (the latter is
the number of different eigenfunctions corresponding to a particular value of
$\lambda$). The corresponding eigenfunctions $u_n$, $n = 1,2,\dots$, form a complete
system in an appropriate Hilbert space. Unlike eigenfunctions of the Dirichlet and
Neumann Laplacian, the first paper about properties of solutions to
(\ref{lap})--(\ref{ort}) had been published by Kuttler only in 1984 (see
\cite{Kut}). Since than, a number of interesting results concerning the so-called
`high spots' of sloshing eigenfunctions has appeared (see the recent review
\cite{Not} aimed at lay readers).

The main result of \cite{Kut} is analogous to Courant's theorem. Namely, if the
eigenfunctions are ordered according to increasing eigenvalues, then the nodes of
the $n$th eigenfunction divide the domain into no more than $n+1$ subdomains. In
view of the additional condition (\ref{ort}), the number of subdomains is $n+1$
instead of $n$ appearing in Courant's theorem. Kuttler's reasoning (a version of
Courant’s original proof), indeed, proves this assertion after omitting the
superfluous reference to the following incorrect lemma.

\vspace{1mm}

\noindent {\it For every eigenfunction of problem $(\ref{lap})$--$(\ref{ort})$ nodal
curves have one end on the free surface $F$ and the other one on the bottom $B$.}

\vspace{1mm}

Counterexamples demonstrating that this lemma is incorrect were constructed twenty
years after publication of \cite{Kut}. They provide various domains $W$ for which
there exists an eigenfunction of problem (\ref{lap})--(\ref{ort}) having a nodal
curve with both ends on $F$. Let us outline the approach applied for this purpose in
\cite{KKM}. The example involves a particular pair velocity potential/stream
function (the latter is a harmonic conjugate to the velocity potential) introduced
in the book \cite{KMV}, \S~4.1.1, namely, \\[-5mm]
\begin{eqnarray}
&& \!\!\!\!\!\!\!\!\!\!\!\!\!\!\!\!\!\!\!\!\!\!\!\! u (x,y) = \int_0^\infty
\frac{\cos k(x-\pi) + \cos k(x+\pi)} {k-\lambda} \, \E {k y} \, \D k , \label{uu} \\
&& \!\!\!\!\!\!\!\!\!\!\!\!\!\!\!\!\!\!\!\!\!\!\!\! v (x,y) = \int_0^\infty
\frac{\sin k(x-\pi) + \sin k(x+\pi)} {\lambda - k} \, \E {k y} \, \D k , \label{vv}
\end{eqnarray} 
where $\lambda = m/2$ and $m$ is odd. Then the numerators in both integrals vanish
at $k = \lambda$, and so they are understood as usual infinite integrals. It is easy
to verify that $u$ and $v$ are conjugate harmonic functions in the half-plane $y <
0$. Moreover, we have that \\[-2mm]
\begin{equation}
u (-x,y) = u (x,y) \quad \mbox{and} \quad v (-x,y) = -v (x,y) , \label{sym}
\end{equation}
which allows us to study the behaviour of nodal curves of these functions only in
the quadrant $\{ x > 0, y < 0 \}$ in view of their symmetry about the $y$-axis.

\begin{figure}[t]
\begin{center}
\SetLabels \L (0.00*0.15) {\small $y$} \\ \L (0.1*-0.01) {\small $x$} \\
\L (0.4691*0.9635) \scalebox{1.07}{\scriptsize $\scriptstyle\bullet$} \\
\endSetLabels
\leavevmode
\strut\AffixLabels{\includegraphics[width=82mm]{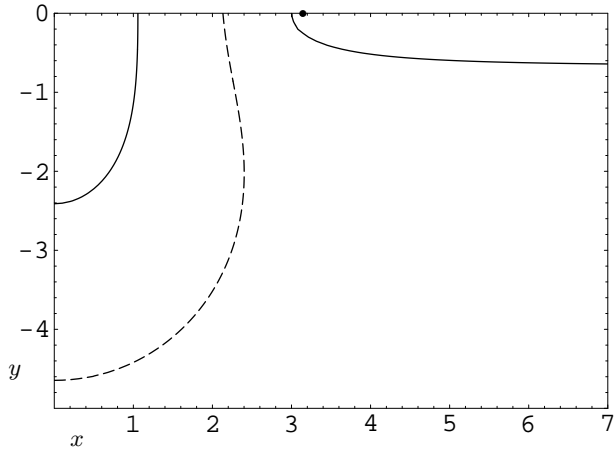}}
\end{center}
\vspace{-5mm} 
\caption{Nodal lines of $u$ (solid lines) and $v$ (dashed line) given
by \eqref{uu} and \eqref{vv}, respectively, with $\nu = 3/2$.} 
\vspace{-4mm}
\end{figure}

In \S~2 of the paper \cite{KKM}, this behaviour is investigated in detail for
$\lambda = 3/2$ and illustrated in Fig.~1, where only the right half of the picture
is shown in view of (\ref{sym}). It is proved that $v$ has a nodal curve which has
its both ends on the $x$-axis (dashed line). This nodal curve serves as $B$ because
the boundary condition (\ref{nc}) is fulfilled on it in view of the Cauchy--Riemann
equations holding for $u$ and $v$. Furthermore, there exists a nodal curve of $u$
(solid line) lying in $W$ defined by the described $B$. Moreover, it has both its
ends on the $x$-axis, thus delivering a counterexample to Kuttler's lemma.

\begin{figure}[t]
\begin{center}
\SetLabels \L (0.03*0.13) {\small $y$} \\ \L (0.13*-0.01) {\small $x$} \\ \L
(0.4942*0.9525) \scalebox{1.07}{\scriptsize $\scriptstyle\bullet$} \\
\endSetLabels
\leavevmode
\strut\kern-4.5mm\AffixLabels{\includegraphics[width=84.15mm]{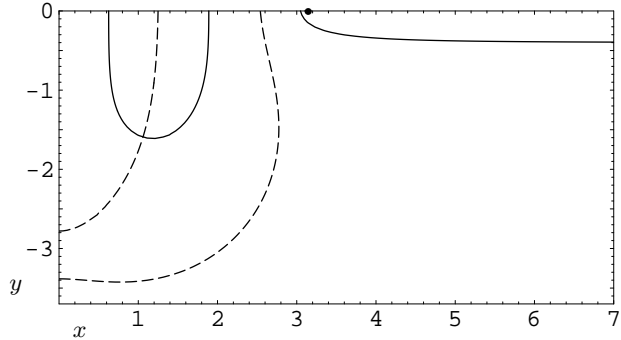}}
\end{center}
\vspace{-5.6mm} \caption{Nodal lines of $u$ (solid lines) and $v$ (dashed lines)
given by \eqref{uu} and \eqref{vv}, respectively, with $\lambda = 5/2$.}
\vspace{-4mm}
\end{figure}

More complicated counterexamples to Kuttler's lemma are obtained numerically for
$\lambda = 5/2$; see Fig.~2, where again only the right half of the picture is
shown. In this case, apart from the $y$-axis there are two nodes of $v$ (dashed
lines and their images in the $y$-axis) and four nodes of $u$ (solid lines and their
images in the $y$-axis). Both finite nodes of $u$ are located within the domain $W$
whose bottom $B$ is given by the whole exterior node of $v$. In another
counterexample, the bottom consists of the right half of this node complemented by
the segment of the $y$-axis.

Besides, taking the interior node of $v$ as the bottom, we see that the nodes of $u$
connect this bottom with the corresponding free surface. Of course, the same is true
for all known cases of the sloshing problem in two and three dimensions for which
separation of variables is possible, thus providing a misleading hint.

\subsection*{On nodal curves of oscillating membranes with fixed boundaries}

The topic of this section is the eigenvalue problem \\[-2mm]
\begin{equation}
u_{xx} + u_{yy} + \lambda u = 0 \ \ \mbox{in} \ D , \quad u = 0 \ \ \mbox{on} \
\partial D , \label{lam}
\end{equation}
where $D$ is a bounded domain in $\RR^2$. Its solutions $(u_n, \lambda_n)$,
$n=1,2,\dots$ (for every $\lambda > 0$ satisfying (\ref{lam}) the number of its
repetitions is equal to its multiplicity) serve for representing pure tones that the
elastic membrane $D$ can produce being fixed along its boundary. As was mentioned
above, along nodal curves an oscillating membrane stays immovable. This is why their
study is important.

In the paper \cite{H}, published after defending his dissertation discussed above,
Herrmann remarked that Courant's theorem admit sharpening for eigenfunctions of
problem (\ref{lam}). Such a refinement appeared in 1956 (see \cite{Pl}), and is
usually referred to as the Pleijel's nodal domain theorem nowadays. Its most
interesting consequence says.

\vspace{1mm}

\noindent {\it The number of subdomains, into which the nodes of the $k$-th
eigenfunction of problem $(\ref{lam})$ divide $D$, is equal to $k$ only for finitely
many values of $k$.}

\vspace{1mm} 

\noindent In the last section of his note, Pleijel writes that ``...it
seems highly probable that the result [\dots] is also true for free membranes'',
that is, when the Dirichlet boundary condition is changed to the Neumann one in
(\ref{lam}). This conjecture was recently proved by Polerovich \cite{Pol} under the
assumption that $\partial D$ is piecewise analytic. The difficulty of this case is
that along with nodal subdomains lying totally in the interior of $D$, there are
subdomains adjacent to $\partial D$, where the Neumann condition is imposed. To the
former subdomains the original technique used by Pleijel and involving the
Faber--Krahn isoperimetric inequality is applicable, whereas the latter ones require
an alternative approach based on an estimate for the number of boundary zeros of
Neumann eigenfunctions.

According to Courant's theorem, the fundamental eigenfunction $u_1$ does not change
sign in $D$, whereas the node of $u_2$ divides $D$ into two subdomains. Both these
cases give the maximal number of subdomains in a trivial way. Less trivial fact
obtained in \cite{Pl} is that only the first, second and fourth eigenfunctions give
the maximal number of subdomains for a square membrane with fixed boundary.

During the past several decades, much attention was paid to the following question.
{\it How does the only node of $u_2$ divide $D$ into two subdomains?} In his widely
cited survey paper \cite{P} published in 1967, Payne conjectured that the nodal
curve of $u_2$ cannot be closed for any domain $D$ (see Conjecture~5 on p.~467 of
his paper)$^0$. It occurred that like Herrmann's theorem this conjecture is only
partly true. The corresponding results are outlined below.

\footnotetext{It is worth mentioning that Yau repeated this question 15 years
later, but only for convex plane domains. May be he expected that it is not true in
its full generality.}
 
Six years later, Payne proved the following theorem confirming his conjecture (see
\cite{P1}).

\vspace{1mm}

\noindent {\it If $D$ is convex in $x$ and symmetric about the $y$-axis, then $u_2$
cannot have an interior closed nodal curve.}

\vspace{1mm}

\noindent Prior to proving this assertion, Payne lists some important facts about
eigenvalues and nodes of eigenfunctions that follow from the theory of elliptic
equations. (In particular, it yields that all solutions of (\ref{lam}) are real
analytic functions in the interior of $D$.) These properties are as follows:

\vspace{1mm}

\noindent (i) If $D'$ is strictly contained in $D$, then the inequality $\lambda_n'
> \lambda_n$ holds for the corresponding eigenvalues.

\noindent (ii) No nodal curve can terminate in $D$.

\noindent (iii) If two nodal curves have a common interior point, then they are
transversal; this also applies when a nodal curve intersects itself.

\vspace{1mm}

Several partial results followed the above Payne's theorem (see references cited in
\cite{Al}) before Melas \cite{M} proved that the conjecture is true for all convex
two-dimensional domains with $C^\infty$ boundary. This happened 25 year after it had
been formulated. Two years later, this result was extended by Alessandrini to the
case of general convex domains in $\RR^2$. Namely, his theorem is as follows (see
\cite{Al}).

\vspace{1mm}

\noindent {\it Let $D$ be a bounded convex domain in the plane. If $u$ is an
eigenfunction corresponding to the second eigenvalue of problem $(\ref{lam})$, then
the nodal curve of $u$ intersects $\partial D$ at exactly two points.}

\vspace{1mm}

\noindent Besides, Payne's conjecture is also true for a class of non-convex planar
domains as was recently shown in \cite{YG}.

Let us turn to results demonstrating that Payne's conjecture is not true for {\it
all} bounded domains to say nothing of unbounded ones. The first counterexample to
the general conjecture in $\RR^2$ belongs to M. Hoffmann-Ostenhof,
T.~Hoffmann-Ostenhof and N. Nadirashvili \cite{HHN} (see also \cite{HHN1}), who
constructed a multiply connected domain such that the nodal set of $u_2$ is disjoint
with $\partial D$.

To describe such a domain we apply non-dimensional variables which is natural from
the physical point of view remembering Arnold's remark about mathematics as a part
of physics. Since the boundary of a domain considered in \cite{HHN} involves two
concentric circumferences (the origin is chosen to be their centre), we take the
radius of the smaller circumference as the length unity. According to \cite{HHN},
the radius of the larger circumference, say, $r \in (1, +\infty)$ must be taken so
that the fundamental eigenvalue of problem (\ref{lam}) in the annulus with interior
and exterior radii equal to 1 and $r$, respectively, lies strictly between the first
and second eigenvalues of problem (\ref{lam}) in the unit circle. These values are
well known being equal to $j_{0,1}^2$ and $j_{1,1}^2$, respectively; here $j_{0,1}
\approx 2.405$ and $j_{1,1} \approx 3.832$ are the least positive zeros of the
Bessel functions $J_0$ and $J_1$, respectively.

The standard separation of variables gives the funda\-mental eigenvalue for the
described annulus. It is equal to $\mu^2$, where $\mu (r)$ is the least positive
root of the following equation \\[-2mm]
\[ J_0 \big(\lambda \big) Y_0 \big(\lambda r\big) - J_0 \big(\lambda r\big) 
Y_0 \big(\lambda \big) = 0 .
\] 
Here $Y_0$ is the zero-order Bessel function of the second kind. Thus, the condition
imposed on $r$ can be written in the form: \\[-2mm]
\begin{equation}
2.405 \approx j_{0,1} < \mu (r) < j_{1,1} \approx 3.832 . \label{bes}
\end{equation}

The existence of $r$ such that (\ref{bes}) is valid is considered by the authors of
\cite{HHN} as an obvious fact and its natural explanation from the physical point of
view is as follows. Since $\mu (r)$ is the frequency of free oscillations of an
annulus with fixed boundary, it monotonically decreases from infinity to zero as the
annulus width $r-1$ increases from zero to infinity, and so inequality (\ref{bes})
holds when $r$ belongs to some intermediate interval. However, it is worth to give a
quantitative evaluation of this interval and this can be easily done with the help
of classical handbooks. The table on p.~204 in \cite{JE} gives that $2$ belongs to
this interval because $\mu (2) \approx 3.123$, whereas Table 9.7 in \cite{AS} shows
that $5/3$ and $5/2$ are out of it because $\mu (5/3) \approx 4.697$ and $\mu (5/2)
\approx 2.073$. More detailed information about the behaviour of $\mu (r)$ one gets
from the graph plotted in figure~110 on p.~204 in \cite{JE}.

The next step is characterised in \cite{HHN1} as ``carving'' $N > 2$ holes in the
circumference separating the unit circle from the annulus in order to obtain a
single multiply connected domain; the angular diameter of each hole is $2 \epsilon$,
where $\epsilon \in (0 , \pi/N)$. Therefore, it is convenient to use polar
coordinates for this purpose: $\rho \geq 0$ and $\theta \in (-\pi, \pi]$ such that
$x = \rho \cos \theta$, $y = \rho \sin \theta$. The boundary of the sought domain
$D_{N,\epsilon}$ is as follows:
\[ \partial D_{N,\epsilon} = \{ \rho = r \} 
\cup \left\{ \rho = 1 , \theta \notin \cup_{k=0}^{N-1} \left( \frac{2 \pi k}{N} -
\epsilon , \frac{2 \pi k}{N} + \epsilon \right) \right\} ,
\]
and so $\partial D_{N,\epsilon}$ consists of $N+1$ (at least three) components. 

Now we are in a position to formulate the main result proven in \cite{HHN} and
\cite{HHN1}. \\[-4mm]

\vspace{1mm}

\noindent {\it Let $r > 1$ be such that inequality $(\ref{bes})$ holds. Then there
exists $N_0 \geq 2$ such that for $N \geq N_0$ and sufficiently small $\epsilon =
\epsilon (N)$ the following assertions are true: {\rm (i)} the 2nd eigenvalue of
problem $(\ref{lam})$ in the domain $D_{N,\epsilon}$ is simple; {\rm (ii)} the nodal
curve of the corresponding eigenfunction $u_2$ is a closed curve in $D_{N,\epsilon}$.}

\vspace{1mm}

\noindent In their proof, the authors use the symmetry of the domain
$D_{N,\epsilon}$. Moreover, they note \\[-6mm]
\begin{quote}
we have not tried to get an explicit bound on the constant $N_0$ [\dots]. This
[\dots] would probably lead to an astronomical number.
\end{quote}
Then they conjecture that no simply connected domain has a closed nodal curve of
$u_2$.

In 2001, Fournais \cite{F} obtained ``a natural higher dimensional generalisation of
the domain'' constructed in \cite{HHN}. Instead of using the symmetry of a domain he
applied an alternative, and in a some sense more direct, approach to ``carving''
evenly distributed holes in the inner sphere in order to obtain the desired
conclusion.

The next step was to consider unbounded domains. In this case, Payne's conjecture
does not hold even for planar domains satisfying conditions used by Payne himself
when proving the conjecture for bounded domains. Namely, the following theorem was
obtained in \cite{FrK}.

\vspace{1mm}

\noindent {\it There exists a simply connected unbounded planar domain which is
convex and symmetric with respect to two orthogonal directions, and for which the
nodal line of a 2nd eigenfunction does not touch the domain's boundary.}

\subsection*{Brief conclusions}

The above examples are taken from a rather narrow area in mathematical physics.
Nevertheless, they clearly show that even incorrect and/or partly correct theorems
and conjectures often lead to better understanding not only of the corresponding
mathematical topic, but, sometimes, a topic in a completely distinct field.

Another conclusion concerns the role of style in Arnold's papers and, especially,
his books. It combines clarity of exposition, mathematical logic, physical intuition
and masterly use of pictures. Therefore, it is not surprising that he is among the
world's most cited authors and No.~1 in Russia according to
http://www.mathnet.ru/php/per son.phtml?\&option\_lang=eng. Every mathematician
would enjoy his papers aimed at a general audience; in particular, \cite{A3} and
\cite{A4}, which show that his English is as excellent as his Russian.
Unfortunately, some translations of his papers leave a lot to be desired (for
example, one finds `knots' instead of `nodes' in \cite{A}; see the top paragraph on
p.~26).

There is a common opinion that Agatha Christie's novels are helpful for learning
English (the author's own experience confirms this). In much the same way, Arnold's
papers and books are helpful for both learning mathematics and learning to write
mathematics.

\vspace{2mm}

\noindent {\bf Acknowledgement.} The author thanks Yakov Eliashberg for photos of V.
Arnold presented in this paper.

\end{document}